\documentclass[
  aps,prb,twocolumn,showpacs,amsmath,amssymb,superscriptaddress
]{revtex4-1}

\makeatletter
\newcommand{\whencolumns}[2]{\preprintsty@sw{#1}{#2}}
\makeatother

\usepackage[english]{babel}
\usepackage{amsmath,amssymb,amsfonts} 	% Typical maths resource package
\usepackage{graphicx}
\usepackage{array,multirow}
\usepackage[utf8]{inputenc}
\usepackage{bm}
\usepackage{xcolor}
\usepackage[normalem]{ulem}
\usepackage{siunitx}
\usepackage{hyperref}
\hypersetup{colorlinks=true,linkcolor=blue,citecolor=blue,urlcolor=blue}
\usepackage{url}
\usepackage{textcomp}
\usepackage{lineno,hyperref}
\usepackage{float}
\usepackage{textcomp} %for n^o
\usepackage{comment}
\usepackage{rotating}
\usepackage[figurename=FIG.]{caption}
\usepackage{subcaption}
\usepackage{color} % only to put comments in the draft
\definecolor{red}{rgb}{1.,0.,0.}
\usepackage{soul}
\usepackage{amsmath}
\usepackage{enumitem}

 \usepackage[
 justification=raggedright]
 {caption}
\usepackage{subcaption}
\usepackage{qcircuit}
\usepackage{braket}

\makeatletter
\newcommand*{\rom}[1]{\expandafter\@slowromancap\romannumeral #1@}
\makeatother

\usepackage{siunitx}
\usepackage{chemformula}
\usepackage{xspace}
 
\newcommand\Harvard{ John A. Paulson School of Engineering and Applied Sciences, Harvard University, Cambridge, MA 02138, USA}
\newcommand\Sapienza{Department of Physics, Sapienza University of Rome, Piazzale Aldo Moro 5, 00185 Rome, Italy }
\newcommand\Bosch{Robert Bosch LLC Research and Technology Center, Watertown, MA 02472, USA}

\DeclareMathAlphabet\mathbfcal{OMS}{cmsy}{b}{n}

\newcommand{\mr}{\mathcal{R}}
\newcommand{\mA}{\mathcal{A}}

\begin{document}

\title{Laser-driven ferroelectricity in SrTiO\textsubscript{3} via quantum fluctuation quenching}

\author{Francesco Libbi}
\affiliation{\Harvard}
\email[Corresponding author. ]{libbi@g.harvard.edu}
\author{Lorenzo Monacelli}
\affiliation{\Sapienza}
\author{Boris Kozinsky}
\affiliation{\Harvard}
\affiliation{\Bosch}

\begin{abstract}
Similar to other perovskites in its family, SrTiO\textsubscript{3} exhibits a significant softening of the ferroelectric mode with decreasing temperature, a behavior that typically heralds the onset of a ferroelectric transition. 
However, this material remains paraelectric down to \SI{0}{\kelvin} due to quantum fluctuations that prevent stabilization of the ferroelectric minimum. This work shows that in the strong out-of-equilibrium regime induced by resonant mid-IR pulses, quantum fluctuations can be suppressed, inducing a ferroelectric transition in \ch{SrTiO3} that is otherwise impossible at equilibrium. The appearance of a metastable state, that is distinct from the conventional ground state, is the first demonstration of how it is possible to leverage and control quantum fluctuations with pulsed light to qualitatively alter the free energy landscape of a quantum system. We predict the conditions and system parameters under which the induced non-equilibrium state can be long-lived and metastable. In providing a quantitative description, based on first principles machine learned potential energy surface, we explain recent experimental observations of light-induced ferroelectric transition in this material. Our results indicate a general nonequilibrium route to light-induced ferroelectric order in oxide perovskites near a ferroelectric instability. 
\end{abstract}

\maketitle

\noindent
Quantum lattice fluctuations play a crucial role in shaping the equilibrium properties of many materials. For example, they stabilize the symmetric phase in $\mathrm{H_3S}$\cite{Errea2016} and $\mathrm{LaH_{10}}$ \cite{Errea2020}, enabling high-temperature superconductivity, and drive the metallization of hydrogen phase III at 380 GPa \cite{Monacelli_Nature_2021}. In perovskites, these fluctuations can stabilize the symmetric paraelectric phase by allowing nuclear wavefunction tunneling through the shallow barrier of the double-well potential energy surface (PES) of the ferroelectric (FE) soft mode \cite{PhysRevMaterials.7.L030801, Ranalli2023,PhysRevResearch.4.033020, PhysRevB.104.L060103}, a phenomenon known as quantum paraelectricity. Notable examples include $\mathrm{KTaO_3}$ \cite{Ranalli2023} and $\mathrm{SrTiO_3}$ (STO) \cite{PhysRevMaterials.7.L030801, PhysRevResearch.4.033020,PhysRevB.104.L060103}. STO, in particular, has been extensively studied due to its frustrated ground state, which arises from the competition between antiferrodistortive (AFD) and FE instabilities \cite{Aschauer_2014}. Below 110 K, the rotation of the AFD mode significantly flattens the PES of the FE soft mode \cite{fechner_quenched_2024},  facilitating quantum tunneling and thereby suppressing ferroelectric order.\\
\noindent
The close proximity of STO to the ferroelectric transition renders it highly sensitive to external perturbations such as epitaxial strain \cite{Haeni2004}, oxygen isotope substitution \cite{PhysRevLett.82.3540, PhysRevLett.96.227602}, or partial cation doping \cite{PhysRev.124.1354}, all of which can stabilize the ferroelectric phase. This inherent sensitivity suggests that dynamic modulation of quantum fluctuations induced by strong THz pulse irradiation may be the driving mechanism behind the ferroelectric transition observed in Refs. \cite{doi:10.1126/science.aaw4911,doi:10.1126/science.aaw4913}.\\
The dramatic response of nuclear quantum fluctuations to nonequilibrium conditions remains largely unexplored. Recent studies have observed that irradiation of cubic STO  with a strong mid-infrared pulse quenches lattice fluctuations in the AFD mode. This mechanism was proposed as a route to potentially stabilizing the ferroelectric phase \cite{fechner_quenched_2024}. However, there is no experimental evidence or atomistic simulations in support of this picture. Furthermore, the dynamics of quantum fluctuations associated with the FE mode, which are expected to play a key role in driving the phase transition, remain entirely unexamined. A more comprehensive investigation is therefore required.\\
In this work, we present the first full-atomistic study of the dynamics of FE quantum fluctuations induced by a strong THz pulse. Our first principles investigation, accelerated by machine learning, predicts and comprehensively explains the light-induced ferroelectric transition reported in Ref. \cite{doi:10.1126/science.aaw4911}. Specifically, we reveal that this transition is driven by the drastic quenching, of FE quantum lattice fluctuations, which ultimately results in the self-trapping of the FE mode in the FE state. Remarkably, we demonstrate that under some conditions it is possible to stabilize this non-equilibrium metastable FE state, making it long-lived and metastable.\\
Other proposed explanations for the light-induced ferroelectric transition have relied on the coupling between phonons and stress \cite{doi:10.1126/science.aaw4911}. This appears to be the mechanism at play when the FE mode is driven directly \cite{PhysRevLett.129.167401, lavoro_sto}, as in the experiments reported in Ref. \cite{doi:10.1126/science.aaw4913}. While certainly interesting, this type of transition can be understood within a semiclassical framework, through modifications of thermodynamic variables such as pressure.
The mechanism we reveal in this work, by contrast, is fundamentally different. It is rooted in the dynamics of nuclear degrees of freedom that have no classical counterpart, representing a genuinely quantum phenomenon. \\
To achieve this, we employ a novel theoretical framework to describe the evolution of the full quantum fluctuation and the nuclear density matrix. This approach extends the recently developed time-dependent self-consistent harmonic approximation (TD-SCHA)\cite{PhysRevB.103.104305,PhysRevB.107.174307,Libbi2025} by incorporating an analytical framework for evaluating quantum ensemble averages. The framework is highly general, encompassing all phonon-phonon interactions up to fourth order, and is rigorously derived from first principles. A detailed derivation of the theory is provided in Ref. \cite{exact_tdscha}, with a summarized account included in the Methods section.

\begin{figure*}[t]
    \centering
    \includegraphics[width=\linewidth]{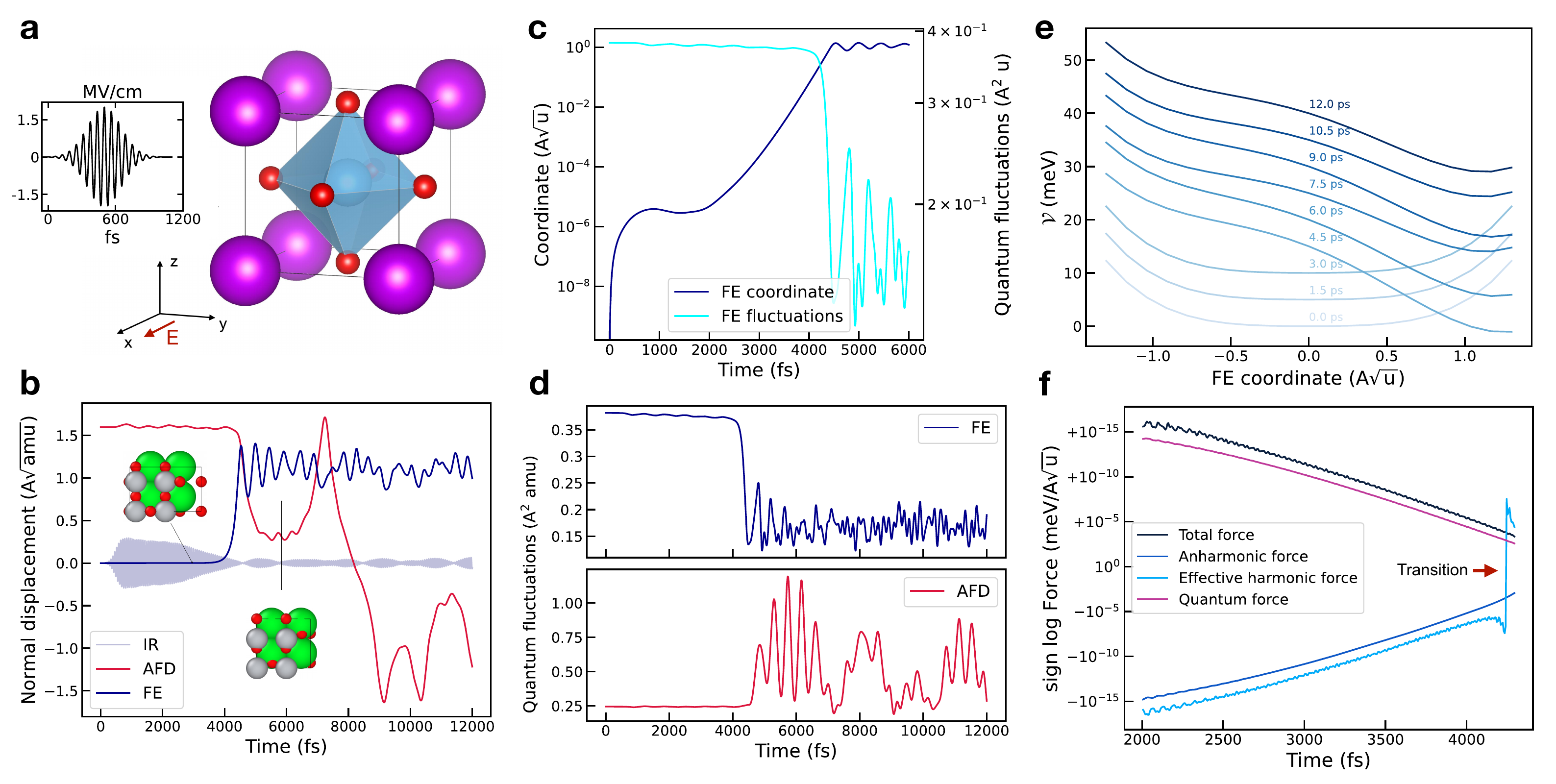}
    \caption{\textbf{a} Schematic representation of the STO unit cell irradiated with a laser pulse. \textbf{b} Displacement of the FE, AFD, and IR modes as a function of time. \textbf{c} Logarithmic plot of the FE displacement and fluctuations. \textbf{d} FE and AFD fluctuations as a function of time. \textbf{e} Nonequilibrium potential energy surface of the FE mode, computed as described in Sec.~V of the SI, shown at several time instants during the dynamics.
    \textbf{f} Signed logarithmic representation of the force components acting on the FE mode. The analysis shows that the quantum force provides the dominant driving contribution to the ferroelectric transition.}
    \label{fig1}
\end{figure*}

\noindent
In our calculations, a 40-atom STO supercell is irradiated with a Gaussian mid-IR pulse of intensity 2000 \text{kV/cm} and frequency 16.0 \text{THz}, polarized along the (1,0,0) direction of the tetragonal STO cell, as represented in \figurename~\ref{fig1}\textcolor{blue}{a}. The pulse starts at approximately 200 fs and ends at around 1000 fs. Initially, the system is in thermodynamic equilibrium in the paraelectric ground state at 0 K.  \\
The pulse selectively excites the high-frequency (15.5 THz) IR-active mode at $\Gamma$, whose dynamics is depicted in \figurename~\ref{fig1}\textcolor{blue}{b}, alongside that of the FE and AFD modes. The pumping field induces pronounced oscillations beyond the linear regime in the resonant IR active vibrational mode, with a peak amplitude of approximately 0.3 $\mathrm{\AA\sqrt{amu}}$. Over time, the amplitude of these oscillations decreases as energy is redistributed to other phonons via anharmonic scattering.
At approximately 4 ps (3.0 ps after the end of the pulse), the FE mode displays a rapid acceleration, ultimately reaching a ferroelectric configuration in which it becomes self-trapped. The transition into the ferroelectric state occurs exponentially in time, as witnessed by the linear displacement of the FE mode from the paraelectric position in  semi-log scale (\figurename~\ref{fig1}\textcolor{blue}{c}). Concurrently, the AFD mode exhibits large chaotic oscillations. 
The FE mode continues to oscillate around the new minimum with larger amplitudes than the pumped mid-IR mode. This is expected, as the elastic energy of a phonon mode scales with the square of its frequency. With a frequency approximately 30 times higher than the FE mode, the IR mode exhibits much smaller oscillations for the same energy. \\
To understand the origin of the laser-induced ferroelectric transition, we analyze the dynamics of the lattice fluctuations $\mathcal{A}_\mu$ associated with the FE mode. These are defined as the quantum-averaged squared displacement of a phonon mode $\mu$ from its mean position,
$\mathcal{A}_{\mu} = \langle (R_{\mu}-\mathcal{R}_{\mu})^2 \rangle$ \cite{PhysRevB.107.174307,PhysRevB.103.104305,Libbi2025}. 
As shown in Fig.~\ref{fig1}\textcolor{blue}{d}, the rapid growth of the FE coordinate around $t \sim 4$~ps is accompanied by a pronounced suppression of the corresponding lattice fluctuations $\mathcal{A}_{\mathrm{FE}}$, which fall below their equilibrium value at $0$~K, $\mathcal{A}_{\mathrm{FE}}^{\mathrm{eq}} \simeq 0.37~\mathrm{\AA^2\,amu}$. In equilibrium, the fluctuation amplitude is given by
$\mathcal{A}_{\mu} = \hbar(2n_{\mu}+1)/2\omega_{\mu}$,
where $n_{\mu}$ is the mode occupation. As the temperature is lowered, $n_{\mu}$ decreases monotonically and vanishes at absolute zero, so that $\mathcal{A}_{\mu}$ reaches its minimum equilibrium value set by zero-point motion,
$\mathcal{A}_{\mu} = \hbar/(2\omega_{\mu})$.
Remarkably, under strong mid-infrared excitation the system enters a nonequilibrium regime in which the FE-mode fluctuations are suppressed below this equilibrium quantum value.  The reduction of $\mathcal{A}_{\mathrm{FE}}$ is accompanied by a concomitant increase in the momentum fluctuations,
$\mathcal{B}_{\mu} = \langle (P_{\mu}-\mathcal{P}_{\mu})^2 \rangle$,
such that the Heisenberg uncertainty relation $\mathcal{A}_{\mu}\mathcal{B}_{\mu} \ge \hbar^2/4$ is satisfied throughout the dynamics (see Sec. I of the Supplementary Information, SI \cite{supplementary}).
As we show below, this pronounced reduction of lattice fluctuations accompanies the rapid growth of the FE coordinate and leads to the localization of the ions in the ferroelectric minimum, thereby stabilizing the ferroelectric phase.\\
We can formulate a minimal model to show that the suppression of fluctuations is a consequence of the exponential growth of ferroelectricity  (\figurename~\ref{fig1}\textcolor{blue}{c}). This model can be derived restricting the full quantum motion to only the ferroelectric dynamics (Sec II of SI):
\begin{equation}\label{eq_Afe}
\begin{aligned}
     \dddot{\tilde\mA}_{FE} + 4\dot{\tilde\mA}_{FE}\Bigl[\omega^2_{FE} + \frac{1}{2}\psi_{FE}(\tilde\mr_{FE}^2+\tilde\mA_{FE})\Bigr]\\+ \psi_{FE}\mA_{FE}(2\dot{\tilde\mr}_{FE}\tilde\mr_{FE}+\dot{\tilde\mA}_{FE}) = 0  
\end{aligned}
\end{equation}
Here, $\tilde{\mA}_{FE} = \mA_{FE} - \mA_{FE}^{eq}$ and $\tilde{\mr}_{FE} = \mr_{FE} - \mr_{FE}^{eq}$ are the deviations of the quantum fluctuations and the FE coordinate from their equilibrium values, respectively, and $\psi_{FE} = \frac{\partial^4 V}{\partial \mr_{FE}^4}$ is the fourth derivative of the PES of the FE mode at the symmetric configuration. Equation~\eqref{eq_Afe} is closely analogous to the model previously introduced in Ref.~\cite{fechner_quenched_2024}, which successfully accounts for the quenching of fluctuations associated with the AFD mode in cubic STO.
Substituting the exponential growth of ferroelectricity $\mr_{FE} \sim \alpha e^{t/\tau}$, we immediately derive the corresponding quenching of quantum fluctuations
$\mA_{FE}(t) \simeq \mA^{eq}_{FE} \Biggl( 1 - \frac{\alpha^2 \tau^2 \psi_{FE}}{4\omega_{FE}^2 \tau^2 + \psi_{FE}\mA^{eq}_{FE}\tau^2 + 4} e^{2t/\tau} \Biggr)$. \\
In a purely classical framework, where the nuclear density is sharply localized ($\mA_{FE}=0$), the nuclei would settle in one of the minima of the double-well PES\cite{PhysRevMaterials.7.L030801, PhysRevResearch.4.033020,lavoro_sto} of the FE mode, resulting in a ferroelectric state. However, quantum fluctuations spread the nuclear density across the potential barrier, enabling tunneling between the wells and stabilizing the paraelectric phase. Consequently, the quenching of the FE fluctuations drives  the system toward the classical regime, abruptly stabilizing the ferroelectric state and leaving it self-trapped in the newly symmetry-broken phase.  This effect is directly reflected in the time evolution of the potential energy $\langle V\rangle$ of the FE mode during the simulation, shown in \figurename~\ref{fig1}\textcolor{blue}{e}. As the lattice fluctuations are quenched, around $t\!\sim\!4$ ps, the minimum of $\langle V\rangle$ shifts toward the ferroelectric configuration, where the system subsequently remains trapped.

\begin{figure*}[t]
    \centering
    \includegraphics[width=\linewidth]{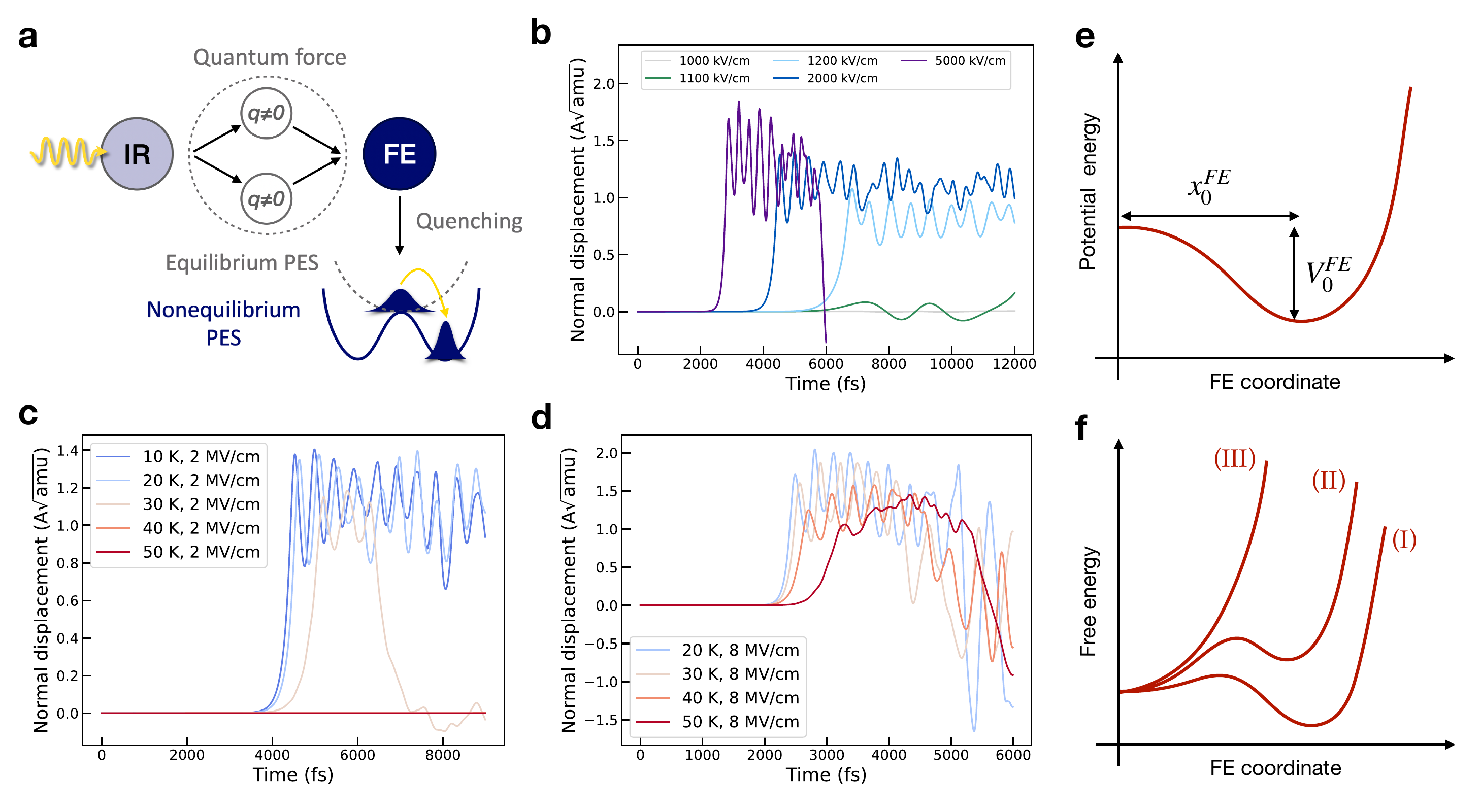}
    \caption{\textbf{a} Schematic representation of the FE transition mechanism.
\textbf{b} Electric-field dependence of the FE transition.
\textbf{c} Temperature dependence of the FE transition for $E = 2000$~kV/cm.
\textbf{d} Temperature dependence of the FE transition for $E = 8000$~kV/cm.
\textbf{e} Schematic representation of the double-well PES of STO.
\textbf{f} Schematic representation of the FES shapes corresponding to regions I, II, and III in panel \textbf{e}.}
    \label{fig2}
\end{figure*}

The growth of ferroelectric order originates from nonequilibrium lattice dynamics that generate an effective dragging force on the FE mode, driving it toward the symmetry-broken state. However, unlike commonly assumed mechanisms \cite{Disa2021}, we find that the resonantly pumped IR mode does not directly drive the FE mode. Instead, it couples to a manifold of finite-momentum phonons, which collectively mediate the response. To make this explicit, we decompose the total force acting on the FE mode, $f_{\mathrm{FE}}$, into three contributions, corresponding to a harmonic (elastic) term, an anharmonic term, and a purely quantum force: 
\begin{align}
 f_{\text{FE}} = -& \overbrace{\sum_\nu\Bigl\langle\frac{\partial^2 V}{\partial \mr_{FE}\partial \mr_\nu} \Bigr\rangle \mr_{\nu}}^\text{harmonic} + \nonumber \\&
-\underbrace{\sum_{\nu\sigma\tau}\frac{1}{3!} \psi_{\nu\sigma\tau\text{FE}} \mr_{\nu} \mr_{\sigma} \mr_{\tau}}_\text{anharmonic}
-\overbrace{\sum_{\nu\sigma}\chi_{\nu\sigma\text{FE}} \mathcal{A}_{\nu\sigma}}^\text{quantum}.
\label{for}
\end{align}
Here, the bracket notation $\langle \cdot \rangle$ denotes the average over the nuclear quantum distribution, the coefficients $\chi_{\mu\nu\sigma}$ and $\psi_{\mu\nu\sigma\tau}$ represent the third- and fourth-order derivatives of the STO potential energy around the symmetric configuration (more details in the Methods section).
The first term on the right-hand side corresponds to the generalized elastic force. The second term represents the classical anharmonic force of a symmetric quartic potential, while the last term accounts for a purely quantum force mediated by the off-diagonal quantum fluctuations $\mathcal{A}_{\nu\sigma}=\braket{\delta R_{\nu}\delta R_{\sigma}}$. \\
\figurename~\ref{fig1}\textcolor{blue}{f} illustrates the signed logarithmic representation (negative values correspond to negative forces) of these three force components projected onto the FE mode. From 2.0 ps to 4.0 ps, the total force acting on the FE mode increases exponentially, as evidenced by the linear trend in the logarithmic plot. During this time interval, the quantum force dominates. By contrast, both the harmonic and anharmonic contributions remain restoring throughout this period, opposing the development of ferroelectric order up to the transition.
This immediately rules out classical dragging mechanisms based on direct anharmonic coupling between the resonantly driven IR mode and the FE mode. Such mechanisms would originate from couplings of the form $\alpha Q_{\mathrm{FE}} Q_{\mathrm{IR}}^{2}$ or $\beta Q_{\mathrm{FE}}^{2} Q_{\mathrm{IR}}^{2}$. The former is forbidden by symmetry, while the latter can induce an instability only for sufficiently strong driving fields, corresponding to unrealistically large IR-mode amplitudes. As shown in Sec.~III of the SI, the dragging condition requires
$A_{\mathrm{IR}} > \omega_{\mathrm{FE}}/\sqrt{|\beta|}\, \simeq 1.2~\mathrm{\AA}\sqrt{\mathrm{amu}}$, 
which is far from satisfied in our simulations, where the maximum IR amplitude reached is $A_{\mathrm{IR}} = 0.4~\mathrm{\AA}\sqrt{\mathrm{amu}}$ (as shown in Fig.~\ref{fig1}\textcolor{blue}{b}). \\

Instead, the activation of the ferroelectricity occurs indirectly through scattering processes that populate off-diagonal components of the quantum fluctuations matrix, as shown by the prominent contribution of the quantum force in \figurename~\ref{fig1}\textcolor{blue}{f}. In particular, due to inversion symmetry in the paraelectric state, the coupling tensor $\chi_{\mu\nu\sigma}$ vanishes when $\mu$, $\nu$, $\sigma$ are zone-center modes, implying that only finite-momentum phonons can mediate this interaction. \\
A direct confirmation of this mechanism is obtained by repeating the simulations while keeping the fluctuation matrix $\mathcal{A}$ frozen at its equilibrium value, in which case no ferroelectric transition occurs. In addition, an analysis of the mode-resolved contributions shows that the quantum force is not dominated by any single phonon pair, but instead arises from a broad distribution of pairs, consistent with a continuum of interacting modes. Both results are reported in Sec.~III of the SI.\\
We can therefore provide a unified picture of the mechanism underlying the ferroelectric transition: (i) A strong laser pulse resonantly excites a high-frequency IR mode (ii) The driven mid-IR mode subsequently decays into pairs of phonons across the Brillouin zone, with wavevectors $\vec q_{\nu}$ and $\vec q_{\sigma}$ satisfying quasi-momentum conservation, $\vec q_{\nu} + \vec q_{\sigma} = \vec G$. The resulting nonequilibrium population of phonon pairs generates an effective \emph{drag force} that drives the system toward the ferroelectric configuration.
(iii) As the ferroelectric distortion grows, lattice fluctuations are progressively quenched, ultimately leading to self-trapping of the system in a ferroelectric state.
A schematic representation of the phenomenon is shown in \figurename~\ref{fig2}\textcolor{blue}{a}.\\

\begin{figure*}
    \centering
    \includegraphics[width=\linewidth]{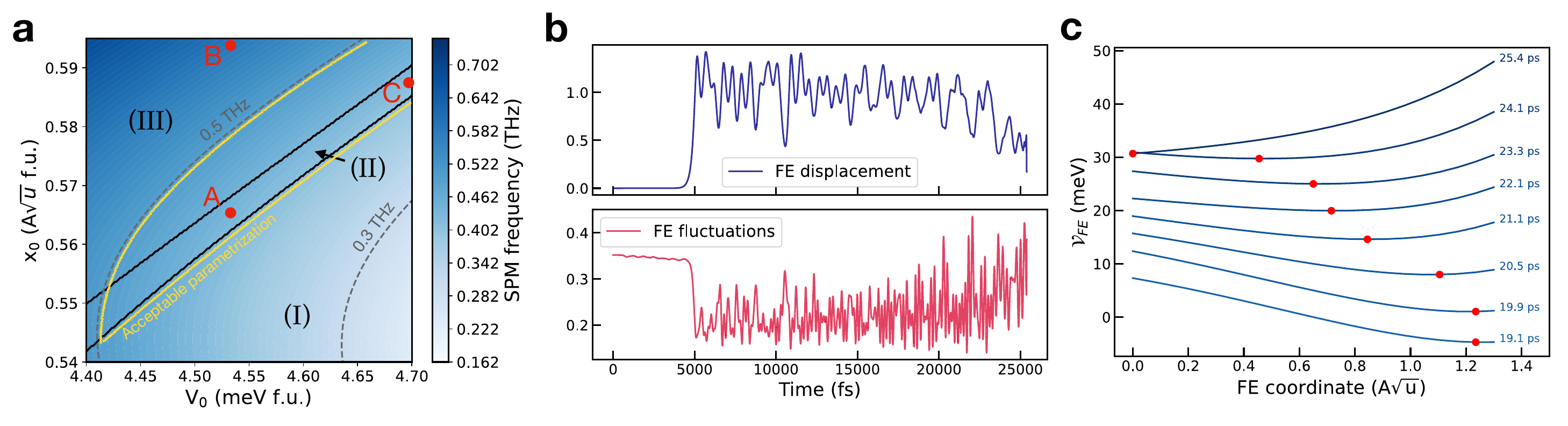}
    \caption{\textbf{a} Map showing the FE-mode frequency and the stability of STO as a function of the potential-energy parameters $(V_0, x_0)$. \textbf{b} Time evolution of the FE-mode displacement and fluctuations calculated for the parameter set corresponding to point B. \textbf{c} Time evolution of the FE-mode potential energy surface during the dynamics at point B, reflecting the increase in lattice fluctuations.}
    \label{fig3}
\end{figure*}

To determine whether there is a threshold in the electric field intensity required to trigger the ferroelectric transition, we repeated our simulations for several field strengths, with the results shown in \figurename~\ref{fig2}\textcolor{blue}{b}.
Interestingly, the transition occurs only for fields greater than 1200 kV/cm.
At 1100 kV/cm, the system exhibits small oscillations, while at 1000 kV/cm, it remains essentially unperturbed in its equilibrium state.
This suggests that the population of finite-momentum phonons—responsible for the quantum driving force—occurs only when the system is driven far from equilibrium, thus highlighting the strongly nonequilibrium nature of the ferroelectric transition.
Moreover, the time at which the transition occurs decreases with increasing electric field strength.
Crucially, when the electric field is very large (5000 kV/cm), the system acquires too much kinetic energy and escapes the ferroelectric minimum, consistent with what is shown in Ref.~\cite{PhysRevLett.129.167401} The precise value of this transition threshold is highly sensitive to the height of the double-well potential energy barrier, as discussed in greater detail later in the manuscript. \\
Photoexcitation with a strong laser pulse inevitably heats STO, raising its temperature above absolute zero, whereas the simulations discussed so far are performed at 0 K. Within the TD-SCHA formalism, energy dissipation occurs through anharmonic phonon--phonon scattering, ultimately leading to thermalization. However, these processes take place on timescales longer than those accessible in our simulations. To explicitly assess the role of thermal fluctuations, we therefore repeat the simulations using initial states prepared at finite temperatures.\\
The results are shown in \figurename~\ref{fig2}\textcolor{blue}{c}. We find that the light-induced ferroelectric transition persists up to temperatures of approximately 30~K. At this temperature, however, the transition becomes transient, lasting about 4~ps before the system relaxes back to the paraelectric phase. At higher temperatures, the transition is suppressed.\\
The temperature threshold can be increased by driving the system with stronger electric fields. As shown in \figurename~\ref{fig2}\textcolor{blue}{d}, for an intensity of 8~MV/cm the transition is also observed at 50~K. In this case, however, the ferroelectric state remains transient, as the system acquires excess kinetic energy and is expected to stabilize only after several oscillation cycles. \\
Does the induced ferroelectricity persist as a metastable state, or is it a transient, dynamical state doomed to fall back into the paraelectric state after a few tens picoseconds? The answer is highly sensitive to the Born-Oppenheimer double-well PES of the FE mode. This double-well function is fully characterized by the height of the potential energy barrier, $V_0^{FE}$, and the position of the wells, $x_0^{FE}$, relative to the symmetric configuration (represented schematically in \figurename~\ref{fig2}\textcolor{blue}{e}). The precise values of these parameters remain uncertain, as experimental techniques are unable to measure them directly, and theoretical calculations based on density functional theory yield results that vary by an order of magnitude depending on the functional employed \cite{PhysRevMaterials.7.L030801}.
To address the presence of metastable ferroelectric states, we computed the free energy landscape over a grid of $x_0^{\mathrm{FE}}$ and $V_0^{\mathrm{FE}}$ values, focusing on the parameter space where the paraelectric FE frequency lies within the experimentally relevant range (0.3–0.5 THz \cite{Yamanaka_2000,doi:10.1143/JPSJ.26.396}) at \SI{0}{\kelvin} (\figurename~\ref{fig3}\textcolor{blue}{a}).  Further details are provided in the Methods section and in Sec. VI of the SI. We identify three distinct regions, whose corresponding energy landscapes are shown in \figurename~\ref{fig2}\textcolor{blue}{f}.
In region (I), located in the bottom right corner of \figurename~\ref{fig3}\textcolor{blue}{a}, the system has a ferroelectric ground state, which contradicts experimental observations. Therefore, the parameters within this region should be excluded from consideration. It is important to emphasize that in this region, the paraelectric state is \textit{metastable}, as evidenced by the positive frequency of the FE mode.
The central stripe (II), delimited by black solid lines, represents parameters where the paraelectric phase is the ground state, coexisting with a metastable ferroelectric phase with slightly higher energy.
In contrast, in the top-left region (III), the ground state is paraelectric, and no metastable ferroelectric phase exists. 
Imposing the additional experimental constraint that the FE-mode frequency lies in the range 0.3–0.5 THz \cite{Yamanaka_2000,doi:10.1143/JPSJ.26.396} further restricts the parameter space, as indicated by the yellow boundary in \figurename~\ref{fig3}\textcolor{blue}{a}.
The resulting physically acceptable parameter set comprises the whole of region (II) and a narrow neighboring part of region (III).
While region (II) appears narrow, it occupies a large part of acceptable parameter space.

The results presented above (\figurename~\ref{fig1}\textcolor{blue}{b-f}) were obtained using a PES in region (II), corresponding to point A in \figurename~\ref{fig3}\textcolor{blue}{a}. To assess the robustness of our conclusions with respect to the choice of the PES, we repeated the simulations using a parametrization placing the FE mode in region (III), corresponding to point B in \figurename~\ref{fig3}\textcolor{blue}{a}. Remarkably, the laser-induced atomic dynamics in region (III), shown in \figurename~\ref{fig3}\textcolor{blue}{b}, also exhibit a ferroelectric transition, qualitatively similar to that observed in region (II). At longer times ($>20$~ps), however, the system relaxes back to the paraelectric configuration, reflecting the purely dynamical nature of the transition in this regime.\\
This result demonstrates that the emergence of a light-induced ferroelectric transition does not rely on a fine-tuned parametrization of the STO potential energy surface, but instead represents a robust outcome of the nonequilibrium dynamics across a broad range of physically relevant PES parameters.
If the experimental values of $V^{FE}_0$ and $x^{FE}_0$ fall within the region (II), this dynamic transition is long-lived. However, if the experimental values of $V^{FE}_0$ and $x^{FE}_0$ are located in region (III), the induced polar order is transient, holding for tens of picoseconds. Further experimental efforts, either direct or indirect, to measure the shape of the FE potential energy surface will ultimately determine which of these scenarios is correct.\\
Interestingly, at long times in the dynamics shown in \figurename~\ref{fig3}\textcolor{blue}{b}, the FE displacement exhibits persistent oscillations whose mean value slowly drifts back toward the symmetric configuration. This behavior arises from the gradual shift of the effective potential minimum toward the paraelectric configuration, driven by the progressive recovery of lattice fluctuations toward thermal equilibrium (\figurename~\ref{fig3}\textcolor{blue}{c}).\\
The parametrization of the PES significantly increases the electric-field threshold required for the system to escape the ferroelectric minimum. As the parameters move toward the upper-right region of zone II, the double-well potential becomes deeper and the ferroelectric minima are located further from the paraelectric configuration, thereby requiring a substantially stronger driving field to destabilize the trapped state.
To quantify this sensitivity, we repeated the simulations for point C in Fig.~\ref{fig2}\textcolor{blue}{a}. As reported Sec. VII of the SI, for this parametrization of the potential energy surface (PES) the escape threshold increases from 5000 kV/cm to 10,000 kV/cm.
Experimentally, Ref.~14 reports a light-induced ferroelectric transition at external fields up to 18,000 kV/cm. This is consistent with the strong dependence of the threshold on the underlying barrier height and is of the same order of magnitude as the values obtained in our simulations.

The ferroelectric transition mechanism proposed in this work does not exclude that alternative pathways, such as the
coupling of the FE mode with strain \cite{PhysRevLett.129.167401,lavoro_sto}, may contribute to ferroelectric ordering. However, as shown in Ref.~\cite{lavoro_sto}, strain alone does not generate metastable ferroelectric states and can therefore account at most for short-lived dynamical responses on picosecond timescales, consistent with experiments in which the FE mode is directly excited \cite{doi:10.1126/science.aaw4913}.\\
By contrast, the mechanism identified here demonstrates the existence of a metastable ferroelectric state and shows that the system can be dynamically driven into it under strong excitation, accounting for the long-lived character of the transition observed by Nova et al. Moreover, our simulations predict the emergence of a long-lived polarization orthogonal to the driving field, in agreement with the same experiment, suggesting that the essential coupling pathways are captured within our framework.
Importantly, while the evidence for a ferroelectric transition in Ref.~\cite{lavoro_sto} is indirect, based on the time-dependent stress profile and the comparison of phonon lifetimes, our work provides a direct first-principles observation of the transition.\\
The simulations are performed on a pristine STO supercell without defects, thereby establishing the transition mechanism in the intrinsic limit. A fully microscopic treatment of disorder would require explicit supercells far larger than those accessible within our TD-SCHA framework and is therefore beyond the scope of the present work. Disorder is expected to act primarily as an additional source of dephasing, affecting quantitative aspects such as lifetimes, while the underlying anharmonic coupling responsible for the instability remains operative. Defects may also modify the proximity of STO to the ferroelectric instability by locally reshaping the potential-energy landscape~\cite{10.1063/1.3139767,PhysRevB.95.035301}. Further theoretical and experimental efforts will be required to clarify the role of defects in the transition. 

The formalism employed in this work treats quantum and thermal fluctuations on equal footing through the $\mathcal{A}$ matrix. While quantum fluctuations dominate at low temperature, the same fluctuation-quenching mechanism acts on thermal fluctuations at finite temperature, providing a general route to light-induced ferroelectricity. As a result, the mechanism identified here is general and applies to materials that are close to a ferroelectric instability and possess the symmetry required for IR-active phonon pumping to couple to the relevant soft mode. This defines a broad class of systems, notably cubic perovskites, which naturally satisfy these symmetry requirements.\\
Since the proximity to the ferroelectric transition can be tuned experimentally through temperature or static strain, allowing full control over the shape of the double-well potential energy surface of the FE mode, our results motivate the exploration of light-driven ferroelectric transitions in cubic perovskites close to the ferroelectric threshold. This may enable the identification of materials hosting more robust metastable ferroelectric states than STO, representing an important step toward ultrafast memory devices based on light-controlled ferroelectricity.

In conclusion, through our fully atomistic quantum simulation of nonequilibrium nuclear motion, we have demonstrated a novel mechanism for ordering matter using intense laser pulses. The highest IR-active phonon, resonantly excited by light, decays into pairs of modes with finite momentum, generating a driving force on the ferroelectric mode. This process suppresses quantum fluctuations,  stabilizing a self-trapped ferroelectric state. \\
Our work provides the first demonstration that quantum fluctuations can be actively controlled with strong THz pulses to reshape the free energy landscape of a quantum system, thereby resolving the previously unexplained origin of the permanent ferroelectric transition reported in Ref. \cite{doi:10.1126/science.aaw4911} and establishing a new framework for dynamical phase control in quantum materials. \\

This research was funded in part by the Swiss National Science Foundation (SNSF, Mobility fellowship P500PT\_217861) and the US National Science Foundation under Grant No. DMR-2119351. Computational resources were provided by the FAS Division of Science Research Computing Group at Harvard University.\\
L.M. acknowledges computational resources from CINECA, proj. IsCb8\_TDSTO

\section{Methods}
Quantum dynamics simulations are performed using the time-dependent self-consistent harmonic approximation (TD-SCHA) \cite{PhysRevB.103.104305,PhysRevB.107.174307} with a recently developed exact integration scheme for ensemble averages \cite{exact_tdscha}. This method eliminates stochastic noise \cite{Libbi2025}, allowing highly accurate simulations, which are essential to investigate the ferroelectric transition given the small energy difference between the paraelectric and ferroelectric phases. Additionally, it provides analytical expressions for the forces as functions of the atomic positions and the quantum fluctuation matrix.
TD-SCHA assumes that the quantum density matrix of a crystal is a Gaussian in the Wigner phase-space \cite{PhysRevB.107.174307, PhysRev.40.749, 10.1063/1.1705323}, parameterized by the average atomic positions $\mr_i$ (also called centroids) and momenta $\mathcal{P}_i$ and the position-position, momentum-momentum and position-momentum fluctuations $\mathcal{A}_{ij}=\braket{\delta R_i \delta R_j}$, $\mathcal{B}_{ij}=\braket{\delta P_i \delta P_j}$, $\Gamma_{ij}=\braket{\delta R_i \delta P_j}$
(see Ref. \cite{Libbi2025} for a complete explanation).
All the variables are mass-rescaled, following the convention introduced in Ref. \cite{PhysRevB.107.174307} (e.g. $\mr_i = \tilde\mr_i\sqrt{m_i}$, where $\tilde\mr_i$ is the position not rescaled by mass). The index $i$ and $j$ label both atoms in the supercell and Cartesian coordinates. To follow the motion of the phonon modes and their quantum fluctuations, it is often convenient to project these variables along the phonon eigenvectors $e_{\mu i}$, where the index $\mu$ represents the phonon branch. This can be done by contracting the indexes of a generic tensor $T_{ij..k}$ with the cartesian indexes of the phonon eigenvector, $T_{\mu\nu...\sigma} = T_{ij...k}\ e_{\mu i}e_{\nu j}...e_{\sigma k}$. Here, the Einstein summation convention is applied, implying summation over repeated indices. For instance, the displacement of mode $\mu$ is computed as $\mr_{\mu} = e_{\mu i}\mr_i$, and is illustrated in \figurename~\ref{fig1}\textcolor{blue}{b} for $\mu$= IR, FE, AFD. The diagonal component of the quantum fluctuations for the mode $\mu$ are calculated as $\mathcal{A}_{\mu\mu}=\mathcal{A}_{ij}e_{\mu i}e_{\mu j}$, and shown in \figurename~\ref{fig1}\textcolor{blue}{d} for $\mu$= FE, AFD.  In general, we adopt the convention that tensors with Greek indices are projected onto the phonon eigenvectors (see, for example, Eq. \ref{for}).
The parameters describing the quantum state of the crystal obey the following set of differential equations.
\begin{equation}\label{tdscha}
    \begin{cases}
    \dot\mr_i = \mathcal{P}_i \\
    \dot{\mathcal{P}}_i = \braket{f_i} \\
    \dot{\mathcal{A}}_{ij} = \Gamma_{ij} + \Gamma_{ji} \\
    \dot{\mathcal{B}}_{ij} = -\kappa_{il} \Gamma_{lj} - \kappa_{jl}\Gamma_{li}\\
    \dot{\Gamma}_{ij} = \mathcal{B}_{ij} - \mathcal{A}_{il}\kappa_{lj}
    \end{cases}
\end{equation}
The term $\braket{f_i}$ corresponds to the ensemble average of the forces on the nuclear quantum distribution, while the term
$
    \kappa_{ij} = \Bigl\langle \frac{\partial V}{\partial \mr_i \partial \mr_j} \Bigr \rangle
$ is the ensemble average of the curvature of the potential energy. \\ In the stochastic implementation of TD-SCHA \cite{Libbi2025}, ensemble averages are calculated using Monte Carlo sampling. While this approach is highly general, it inevitably introduces stochastic noise. Alternatively, by expanding the potential energy up to the fourth-order term, these ensemble averages can be computed analytically \cite{exact_tdscha}, ensuring very high accuracy in time integration. The analytic expression for the ensemble averages of forces coincides with Eq. \ref{for} of the main text, while that for the curvature is
\begin{equation}
    \kappa_{ij} = \phi_{ij} + \chi_{ijk}\mr_k + \frac{1}{2}\psi_{ijkl}(\mr_k\mr_l+A_{kl})\ .
\end{equation}
The Taylor expansion of the potential energy surface is applied to the symmetric configuration of tetragonal STO, which corresponds to the centrosymmetric $P4/mmm$ phase.
In this context, the second, third, and fourth derivatives of the potential energy surface are represented by the tensors $\phi_{ij}$, $\chi_{ijk}$, and $\psi_{ijkl}$, respectively. These tensors are computed using a finite difference scheme that employs a machine-learning interatomic potential (MLIP) for STO. Details regarding the training of the MLIP can be found in Refs. \cite{lavoro_sto, Libbi2025, Vandermause2020, Xie2021}. 
It is important to emphasize that, in this exact formalism, the coordinate $\mr_i$ represents the displacement of atom $i$ relative to the reference structure used for the Taylor expansion.  \\
The lattice parameters for STO are determined through cell relaxation at 0 K, performed using SCHA in its stochastic implementation \cite{Monacelli_2021, PhysRevB.96.014111}  in conjunction with the aforementioned MLIP.\\
Dynamic simulations are carried out on a system initially equilibrated at 0 K. Equilibrium positions and fluctuations are obtained through SCHA relaxation. The coupling between laser light and nuclear coordinates is described by the Born effective charges, which are taken from Ref. \cite{lavoro_sto, Libbi2025} and evaluated at the equilibrium structure. These charges are assumed to remain constant throughout the simulation.\\
The pulse applied to the system has Gaussian shape, described by the function $E(t) = A \cos(\omega t)e^{-\frac{t^2}{2\sigma^2}}$, with $\mathrm{A=2\ MV/cm}$, $\mathrm{\omega = 16.0\ THz}$ and $\mathrm{\sigma=150\ fs}$. The electric field is applied along the $x$ direction of the tetragonal cell, with a small tilt of $0.1^\circ$ about the $y$ axis.  We employ an explicit Runge-Kutta method of order 5(4) for the time integration. \\
To assess the role of thermal fluctuations, we performed simulations starting from finite-temperature initial states (Fig.~\ref{fig2}\textcolor{blue}{c,d}).
Because STO is a wide-bandgap insulator (\(E_g \gtrsim \SI{3.2}{\electronvolt}\)) and the highest phonon energies are \(\lesssim \SI{0.1}{\electronvolt}\), direct dissipation of the driven phonon population into electronic excitations is expected to be negligible under the conditions considered here. \\
The parameters $x_0^\mu$ and $V_0^\mu$ of the potential energy surface for mode $\mu$ are related to the diagonal components of the second- and fourth-order derivative tensors by the expressions $x_0^\mu = \sqrt{-\frac{6\phi_{\mu\mu}}{\psi_{\mu\mu\mu\mu}}}$ and $V_{0}^\mu = \frac{3 \phi_{\mu\mu}^2}{2\psi_{\mu\mu\mu\mu}}$.
By acting on these diagonal components for the mode $\mu = \text{FE}$, it is possible to impose the desired values of $x_0^{FE}$ and $V_0^{FE}$. The detailed strategy for this modification is outlined in Ref. \cite{exact_tdscha}.
Notably, since the energy barrier of STO is on the order of a few meV, the resulting adjustments to the force constants are minimal, ensuring that all other phonon modes and their couplings remain unaffected.\\
The colormap in \figurename~\ref{fig2}\textcolor{blue}{d} is obtained by performing SCHA relaxations on a uniform grid of values $(x_0^{FE}, V_0^{FE})$, followed by the calculation of the free energy Hessian. The eigenvalues of this Hessian correspond to the physical frequencies of STO \cite{PhysRevB.96.014111}. For each grid point, the FES along the FE coordinate is calculated by performing SCHA minimizations with respect to the quantum fluctuations matrix only for different displacements of the FE mode.

%\appendix


%\include{Supplementary}
%\bibliographystyle{ieeetr}
%\bibliography{main.bib}
\end{document}